# First observation of liquid-xenon proportional electroluminescence in THGEM holes

L. Arazi,[a,*] A. E. C. Coimbra,[b] R. Itay,[a] H. Landsman,[a] L. Levinson,[a] M. L. Rappaport,[a] D. Vartsky [a,†] and A. Breskin[a]

[a] *Weizmann Institute of Science,*
   *76100, Rehovot, Israel*

[b] *University of Coimbra,*
   *Coimbra, Portugal*
   *E-mail:* lior.arazi@weizmann.ac.il

ABSTRACT: Radiation-induced proportional-electroluminescence UV signals, emitted from the holes of a Thick Gas Electron Multiplier (THGEM) electrode immersed in liquid xenon, were recorded with a PMT for the first time. Significant photon yields were observed using a 0.4 mm thick electrode with 0.3 mm diameter holes; at 2 kV across the THGEM the photon yield was estimated to be ~600 UV photons/electron over $4\pi$. This may pave the way towards the realization of novel single-phase noble-liquid radiation detectors incorporating liquid hole-multipliers (LHM); their concept is presented.

---

[*] Corresponding author
[†] On leave from Soreq NRC, Yavneh, Israel

# Contents

## 1. Introduction

Noble-liquid detectors have been employed in numerous fields for several decades. Examples are in particle-physics calorimetry, gamma imaging in astronomy and medicine, astro-particle physics, etc. Among the most significant applications are neutrino physics and dark-matter (DM) searches. Detailed reviews of present noble-liquid detector techniques and applications can be found in [1-3].

Some applications rely on the detection of primary scintillation-photons resulting from radiation-induced excitation of the noble liquid, in so-called *single-phase* detectors. The detection is usually performed with dedicated vacuum photomultipliers (PMTs) immersed in the cryogenic noble liquid (e.g., XMASS liquid xenon dark-matter experiment [4]); it relies on measuring the scintillation light emitted promptly at the site of interaction. In some cases, the radiation-induced charges are collected within the liquid as well, to provide localization [5]. More complex are *dual-phase* (liquid and gas) detectors, e.g., those developed for detecting Weakly Interacting Massive Particles (WIMPs) – a hypothetical dark-matter candidate. Their detection would consist of the observation of low energy (keV-scale) nuclear recoils resulting from their rare scattering events in a noble-liquid Time Projection Chamber (TPC) detector. Liquid argon (LAr), e.g., in ArDM [6] and liquid xenon (LXe), e.g., in XENON100 [7], LUX [8] and others, are the preferred media in such dual-phase detectors, allowing for the construction of large detection volumes.

Dual-phase TPC detectors record two signals: the primary prompt scintillation light (S1) within the liquid and a secondary delayed signal (S2) generated by the recoil-induced ionization electrons, liberated at the site of interaction, as they pass through the gas phase after extraction from the liquid. S2 can be a result of electroluminescence in the saturated vapor phase above the liquid [7] or of a multiplied charge [6] in this gas gap. The distinct difference between the secondary-to-primary signal ratios (S2/S1) for nuclear and electron recoils (from gamma background) is the key to the efficient distinction between the two [7].

Present-day noble-liquid detectors and others under development (e.g., XENON1T [9]) employ large, costly PMT arrays that fulfill the strict requirements of current experiments: they withstand cryogenic conditions and have low natural radioactivity and high single-photon detection efficiency at the relevant LXe UV-emission wavelength (178 nm). However, their high cost calls for novel, more affordable solutions for future generations of multi-ton detectors, e.g., DARWIN [10].

One concept, in advanced R&D, relies on photon detection in single-phase and dual-phase TPCs with large-area gas-avalanche photomultipliers (GPMs) [11-13]. These are expected to have lower cost, flat geometry, high pixilation and low natural radioactivity. A typical GPM consists of cascaded hole avalanche multipliers (e.g., Gas Electron Multipliers (GEMs) [14] or Thick Gas Electron Multipliers (THGEMs) [15]), or hybrid combinations of hole and mesh multipliers [16, 17], with the first element being coated with a CsI UV-photocathode [18]. Both the THGEM-GPM and the hybrid-GPM showed high gains in combination with an LXe-TPC (within a medical Compton-camera project) [16, 17]. The GPM concept has been under intense R&D at the Weizmann Institute, within the framework of DARWIN [10], for future multi-ton DM detectors; it has been recently considered by the PANDA-X projected DM experiment, also in a single-phase configuration [19, 20]. A single-phase LXe detector coupled to a GPM is under development for combined fast-neutron and gamma imaging [21].

Besides costs, the increase of the dimensions of dual-phase multi-ton devices might not be technically simple; particularly because of the necessity to extract electrons from liquid into gas through large, very flat and parallel mesh-electrodes, keeping constant temperature (and pressure) across the detector. Therefore, a novel concept was recently proposed that would permit the efficient recording of both low-energy recoil-induced scintillation-light and ionization-electron signals in large-volume single-phase noble-liquid detectors [22]. It is based on the detection of both UV-photons and ionization electrons by Liquid Hole-Multipliers (LHMs). These consist of a combination of cascaded GEMs, THGEMs, Micro-Hole & Strip Plates (MHSPs) [23] or other dedicated electrodes, coated with CsI photocathodes. CsI photocathodes have high quantum efficiency for both LAr and LXe emission wavelengths [18]; with CsI immersed in liquid xenon, a quantum efficiency of ~23% was measured at 178 nm under a 10 kV/cm field [24, 25]. The proposed detection process is schematically shown in Figure 1. Electrons photo-induced from CsI by primary scintillation and event-correlated drifting ionization electrons are collected within the liquid by a strong dipole field into the multiplier's amplification holes. They generate electroluminescence photons as they drift in the high electric field within the holes and, depending on the field strength, may also undergo some modest charge multiplication. Forward-emitted UV-photons (shown in Figure 1) impinge on the photocathode of the next element in the cascade, inducing additional photoelectrons that generate UV photons within its holes; the process can be repeated until a detectable photon (or charge) signal is reached (even without charge multiplication). This process is similar to a photon-assisted electron multiplication method evaluated for avalanche-ion blocking with cascaded multipliers [26, 27]. If the total amplification is large, the final charge signals can be collected on readout pads immersed in the liquid; alternatively, photon flashes from the last element can be recorded by photon detectors deployed within the liquid – e.g. large-area GPMs. To reduce possible photon-feedback effects generated by electroluminescence (e.g., secondary electrons from electroluminescence photons emitted backwards into the TPC) the holes in the electrodes of the cascaded stages could be staggered. More details on this newly proposed concept can be found in [22]. Note that another, different, single-phase spherical-TPC noble-liquid detector concept was proposed (comprising a GEM immersed in the liquid); it has not yet been materialized [28]).

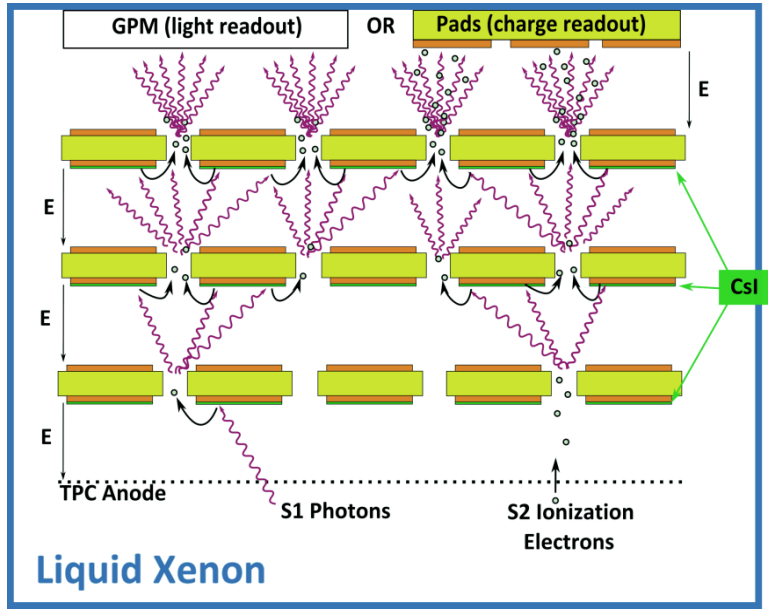

**Figure 1**: The proposed Liquid Hole-Multiplier (LHM) principle, here shown with 3 THGEM electrodes. Radiation-induced S1 UV-photons impinge on the first CsI-coated THGEM electrode; they eject photoelectrons which are pulled into the holes, leading to electroluminescence. The process of light amplification is repeated in the next stages. Similarly, drifting radiation-induced S2 ionization electrons are focused into the holes and follow the same amplification path. The resulting S1 and S2 light signals are recorded optically by an immersed Gaseous Photomultiplier (GPM); if of sufficiently large yield, photon-induced electrons from the final stage, forming the S1 and S2 signals, can be simply collected on readout pads.

Moderate charge multiplication (a few hundred) [29, 30] and electroluminescence (~100 photons/electron at a charge gain of ~50) [31] were observed in the past in LXe, on few-micron diameter anode wires immersed in the liquid, as also reviewed in [3, 32]. Proportional electroluminescence in LXe was recently characterized for a 10 μm diameter gold plated tungsten wire, within an effort to develop a single-phase detector with S2 UV-photon emission from wires [33]. Preliminary values of the measured onset of proportional electroluminescence in [33] are in agreement with those observed in [31]. In LAr, charge gains of ~100 were reached only on sharp (0.25 μm radius) tips [34]. Electroluminescence in LAr (estimated yields reaching 500 photons/electron over 4π), without charge multiplication, was recently demonstrated in THGEM holes, at rather moderate fields [35].

In this work we demonstrate for the first time the successful proportional electroluminescence process occurring within the holes of a THGEM electrode immersed in LXe, with both gamma photons and alpha particles. We provide an estimate of the photon yields – namely, the number of photons emitted per electron drifting along the THGEM hole, as a function of the applied voltage. We discuss our first results and suggest further research steps necessary for the realization of the LHM concept and its potential application in future, large scale, rare-event detection experiments.

## 2. Experimental setup and methodology

### 2.1 Liquid xenon setup

The LHM investigations were carried out in WILiX, the Weizmann Institute Liquid Xenon system, comprising a super-insulated cryostat, a gas handling and purification system, and a 250-liter xenon gas storage/recovery reservoir.

The cryostat, shown schematically in Figure 2, comprises a 510 mm diameter × 390 mm high outer vacuum chamber (OVC) and a 360 mm diameter × 140 mm high stainless steel inner vacuum chamber (IVC), containing the LXe detector. The relatively large dimensions of the cryostat were chosen to allow for the development of large-area *single-* and *dual-phase* detectors. In the present experiments, the investigated detector elements were placed in a central 86 mm diameter region, with the unused IVC volume filled with a Teflon block.

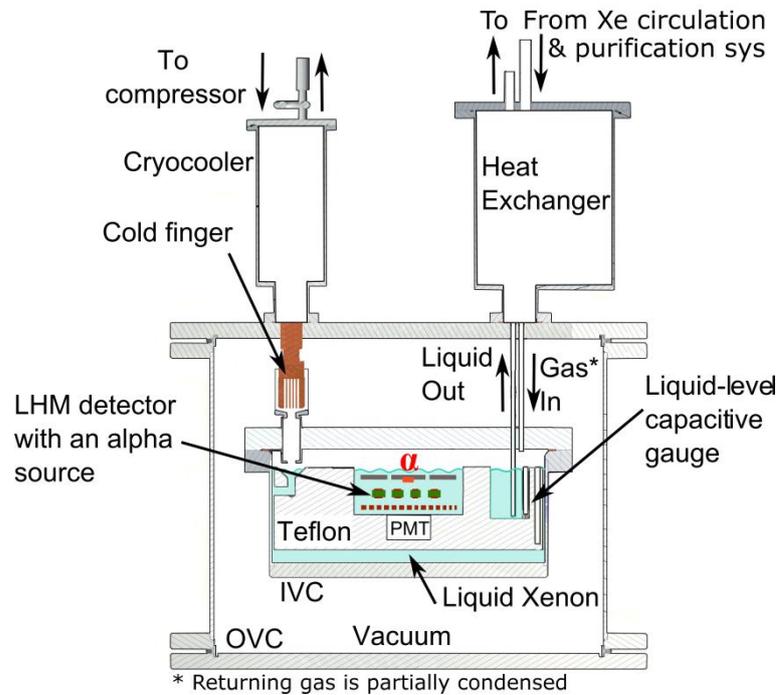

**Figure 2**: Schematic drawing of the WILiX liquid-xenon setup, with the LHM assembly (not to scale) at its center. The LHM configurations are shown in Figure 3. See text for details.

Xenon gas condenses on the finned-end of a temperature-controlled copper cold finger inside the IVC that is thermally linked to the cold stage of a Brooks Automation PCC J-T cryocooler (maximum cooling power 28 W at 128 K). A Cryo-con Model 24C temperature controller is used to control the temperature of the cold finger, closing the control loop with a Pt100 temperature sensor and a 50 W cartridge heater. LXe droplets forming on the cold finger are funneled towards the IVC wall. The liquid fills the central region of the Teflon block (the detector space) from below and overflows into a small reservoir from which it is siphoned out through a commercial parallel plate heat exchanger (GEA model GBS100M). Xenon gas flowing out of the heat exchanger passes through a mass-flow controller (MKS model Mass-flow 1479A) to a double-diaphragm recirculation pump (KNF model N143SV.12E), through a SAES MonoTorr hot getter model PS4-MT3 and then returns to the IVC through the heat exchanger in which ~95% is re-liquefied. Typical flow rates are ~3-5 standard liters per minute (slpm) – the nominal flow rate of the getter. The liquid level inside the IVC is determined by measuring the capacitance of two cylindrical capacitors: one (94 mm long) close to the IVC inner wall and the other (54 mm long) in the small overflow reservoir.

## 2.2 LHM setup

The LHM setup, shown schematically in Figure 3, used a circular THGEM electrode with a 34 mm diameter active area. The THGEM was immersed in the liquid between a bottom electroformed copper mesh with 90% transparency (Precision Eforming, MC17), and a perforated top plate holding an alpha-particle source. Scintillation-light signals were recorded by a 2" diameter vacuum PMT (Hamamatsu model R6041-06), whose QE at 178 nm was measured to be ~28%. The bottom mesh was located 5 mm above the PMT window and 2.5 mm below the THGEM. The top plate held the alpha source 5 mm above the THGEM. The gold-plated FR4 THGEM electrode had the following geometry: thickness $t$=0.4 mm; hole-diameter $d$=0.3 mm; holes pitch $a$=1 mm; etched rim-size around holes $h$=0.1 mm. Two modes of irradiation were investigated: (a) external irradiation with 662 keV gamma photons from an 8 µCi $^{137}$Cs source placed below the OVC, and (b) internal irradiation with ~4 MeV alpha particles (and 60 keV gammas) from a non-spectroscopic 0.4 µCi Au-coated $^{241}$Am source. The PMT was operated at 850-1000 V, recording the primary scintillation signals (S1) from the drift gap and the electroluminescence signals (S2) from the holes. The PMT signals were fed directly (without amplification) into a digital oscilloscope (Tektronics 5054B or Agilent DSO9064A), and processed off-line with dedicated software tools developed for this purpose.

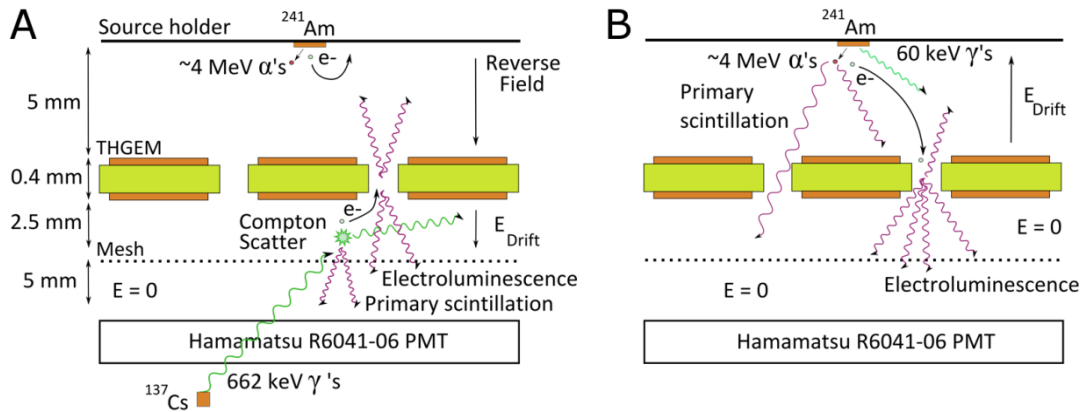

**Figure 3:** Schematic drawing of the LHM setup (not to scale). (A) Irradiation with 662 keV gammas from an external source located below the cryostat; (B) Irradiation with an internal $^{241}$Am alpha source. A reverse field is applied above the THGEM in (A) to prevent alpha-particle induced ionization electrons from reaching the THGEM. See text for additional details on the modes of operation.

The system was operated with the cold-finger temperature set to 162.5 K. The IVC pressure was 1.3-1.5 bar throughout the experiment, corresponding to a temperature of 173-175 K of the liquid. The liquid level was above the plate holding the alpha source, allowing for the extraction of liquid rather than gas through the heat exchanger to the gas recirculation system. The flow rate was 4-5 slpm.

When operating with an external gamma source, a drift field $E_{drift}$ = 1 kV/cm was applied between the bottom mesh and the THGEM, with a reverse field between the THGEM and the alpha-source plate (Figure 3A), to prevent alpha-induced ionization electrons from reaching the THGEM holes. Thus, gamma photons absorbed or Compton-scattered in the gap between the bottom mesh and the THGEM induced primary-scintillation (S1) signals at the interaction point in the liquid; these were followed by electroluminescence pulses (S2) originating from the ionization electrons as they drifted up through the THGEM holes. The time difference between the two signals depended on the depth of interaction (relative to the THGEM bottom); the duration of S2 signals depended on the length and inclination of the Compton- or photo-electron tracks (typically <1 mm total length). When operating the LHM with alpha particles, a drift field of 0.1-3 kV/cm was applied between the THGEM top and the source plate, typically with zero-field in the lower gap (Figure 3B). In this case, all of the events occurred essentially at the same distance (5 mm) from the THGEM top (the range of 4 MeV alpha particles in LXe, as calculated by SRIM [36], is 27 µm). Thus the time difference between S1 and S2 signals was the same for all events, depending only on the magnitude of the drift field.

Measurements performed using the external $^{137}$Cs gamma source were mostly qualitative in nature because of the large variance in energy and depth of the primary interaction. Thus, this mode of operation was used primarily to demonstrate the appearance of S1+S2 signal pairs for electron recoils and show their variability due to the different structure of the events. The second mode of operation, with alpha particles, allowed for several quantitative measurements. These included: recording the spectrum of primary scintillation – later needed for S2-signal normalization; measuring the electron drift velocity (as deduced from the time difference between S1 and S2) and its dependence on the drift field and finally - measuring the spectrum of S2 pulses as a function of the THGEM voltage and the drift field.

The results presented below were obtained after about two weeks of xenon purification, at a flow of 4-5 slpm. The liquid purity is unknown, as electron lifetime could not be measured with the small drift gaps employed; however, since a large number of ionization electrons were able to reach the THGEM from the alpha track, we estimate that the lifetime is at least comparable to the drift time across 5 mm (i.e., on the order of ~1-2 µs or more). This value suggests an impurity level on the ppm scale or lower [1, 37].

## 3. Results

### 3.1 Irradiation with 662 keV gammas

Figure 4 shows four examples of S1 and S2 signals obtained for external irradiation with 662 keV gammas (Figure 3A). The voltage across the THGEM was $V_{THGEM}$ = 3 kV, the drift field was $E_{drift}$ = 1 kV/cm and the reverse field above the THGEM was 2 kV/cm. In this configuration the maximum stable THGEM voltage with gammas was $V_{THGEM}$ = 3.2 kV; at higher voltages occasional discharges appeared between the THGEM top and bottom. The PMT was biased at 1000 V. A large fraction of the S1 pulses were of large amplitude (>100 mV), resulting from the large, unobstructed, solid angle subtended by the PMT relative to the interaction point below the THGEM. The typical width of S1 pulses was ~50 ns, with a gradual decay over several hundred ns. In most cases the Compton-scattered gammas deposited a small amount of energy in the liquid. In these cases, the number of ionization electrons was small, and the S2 signal was of low amplitude, comprising a series of low-amplitude pulses, typically within ~1.5 μs of the sharp, large-amplitude S1 pulse. Quite often, these small S2 pulses were difficult to separate from the decaying tail of S1, which extended up to ~0.5 μs after the initial rise of the pulse. Figure 4A shows an example of a low-amplitude S2 pulse. Note that this case may actually be a double-scatter event, with the first, low-amplitude S2 signal mixed with the decaying tail of S1. Figure 4B shows an event where a large S2 signal begins immediately after S1 – indicating that the event occurred either within a THGEM hole or immediately below it. Figure 4C shows an example where S1 and S2 are clearly separated; the time difference between them (~0.8 μs) suggests that the event occurred ~2 mm below the THGEM. Lastly, Figure 4D shows an example where a single event contains two clear S2 signals (S2, S2') within 1.5 μs of S1, apparently followed by a third S2 signal (S2'') ~1 μs later – indicating a multiple-scatter event. Note that the typical duration of S2 is ~0.5-1 μs in all cases, with specific variations arising from differences in the length and orientation of the Compton- (or photo-) electron tracks. Since the time required for a drifting electron to cross the THGEM hole is ~0.2-0.3 μs (~0.5 mm with a drift velocity of ~2 mm/μs [38]), the pulse duration is likely dominated by the size of the drifting electron cloud with some enhancement resulting from the decay time of the scintillation process.

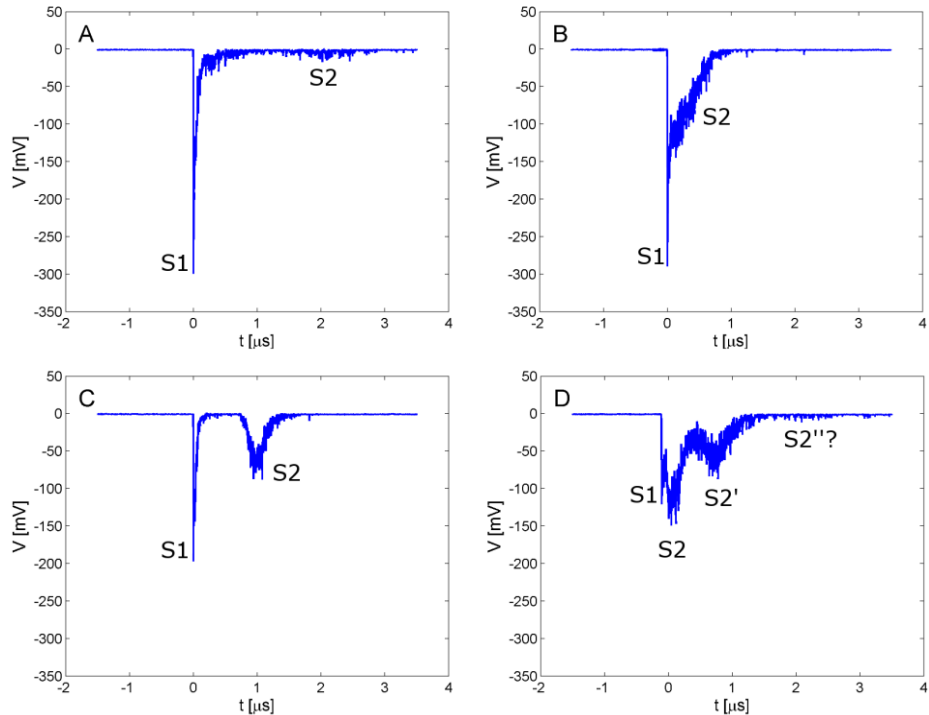

**Figure 4:** Examples of S1 and S2 signals obtained in LXe with 662 keV gammas emitted by an external $^{137}$Cs source (setup of Figure 3A). (A) Low amplitude S2 resulting from a small deposition of energy (likely in a small-angle Compton scatter); (B) An event occurring inside a THGEM hole or immediately below it; (C) Single scatter event, occurring ~2 mm below the THGEM; (D) Multiple scatter event. In all cases $V_{THGEM}$ = 3 kV and $E_{drift}$ = 1 kV/cm.

## 3.2 Irradiation with alpha particles

Unlike the operation with the $^{137}$Cs source, in which the gamma interactions leading to S1+S2 signal pairs occurred between the THGEM and bottom mesh, all of the events associated with the alpha source (both alpha and 60 keV gamma emissions), occurred above the THGEM (Figure 3B). For gammas interacting below the THGEM, the probability for a primary-scintillation photon to reach the PMT was 0.3-0.45. On the other hand, the probability for S1 photons emitted isotropically from the alpha-particle interaction point to reach the PMT (taking into account the solid-angle subtended by the PMT relative to the $^{241}$Am source and the geometry of the THGEM holes), was calculated to be $2.2\times10^{-3}$. For a typical alpha-particle energy of 4 MeV, and assuming that 18 eV are required on average to produce a single UV photon [1], the number of photons emitted on average per alpha particle interaction was $\sim2.2\times10^{5}$. Thus, about 500 photons reached the PMT on average per alpha particle emitted into the liquid. With the PMT operated at 850 V, the resulting S1 signals with alpha particles were rather small (compared to those obtained with 662 keV gammas) - of the order of a few mV. Figure 5 shows a histogram of the integrated pulse-area (in units of mV·µs) of the alpha-induced S1 pulses (with $V_{PMT}$= 850 V), for a zero drift field (setup of Figure 3B). We focus here on the integrated pulse-area rather than its amplitude, because – not having used a charge sensitive preamplifier - it is a more reliable measure of the total number of photons reaching the PMT. It also enables estimating the number of photons reaching the PMT in S2 pulses by taking the ratio of pulse-areas S2/S1. (Note that since S2 has a different pulse shape than S1, the amplitude ratio would not be useful for this purpose.)

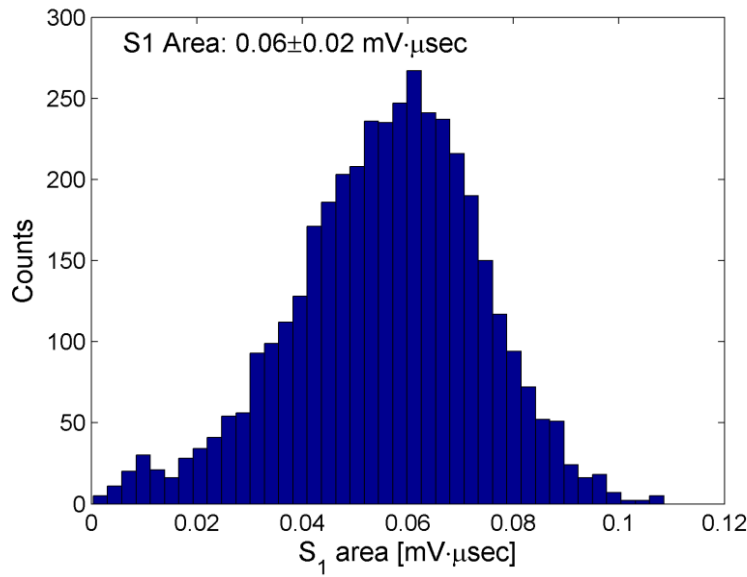

**Figure 5:** Histogram of alpha-particle induced S1 pulse-areas in LXe, recorded with a PMT voltage of 850 V for $E_{drift}$ = 0 (setup of Figure 3B).

With alpha particles, faint occasional S2 signals began to appear at $V_{THGEM} \approx$ 300 V, becoming clearly distinct from the noise at $V_{THGEM} \approx$ 500 V. S2 signals increased in magnitude with increasing $V_{THGEM}$. For $E_{drift}$ = 1 kV/cm, the time difference between S1 and the initial rise of S2 was 2.3 µs (corresponding to a drift velocity, $v_d$ = 2.2 mm/µs). The maximum stable

THGEM voltage in these conditions was 2.6 kV, at which occasional discharges started to appear between the THGEM top and bottom faces. Figure 6 shows samples of S1+S2 signal pairs for alpha particles, at increasing THGEM voltage. In all cases $E_{drift}$ = 1 kV/cm and the field between the THGEM and bottom mesh is zero. The PMT voltage was kept at 850 V. The typical width of the S2 signals (at half maximum) was 0.2-0.3 µs, with a total width of ~0.5-0.6 µs. Unlike with 662 keV gammas, there was very little variation in the S2 pulse shape. This can be attributed to the small size of the electron cloud formed along the ~30 µm alpha-particle track and to the small value of the longitudinal diffusion coefficient of electrons in LXe: $D_L \approx 10$ cm$^2$/s [1], resulting in a longitudinal spread of the order of $\sqrt{D_L t_d} \approx 45$ µm (for a diffusion time of 2 µs).

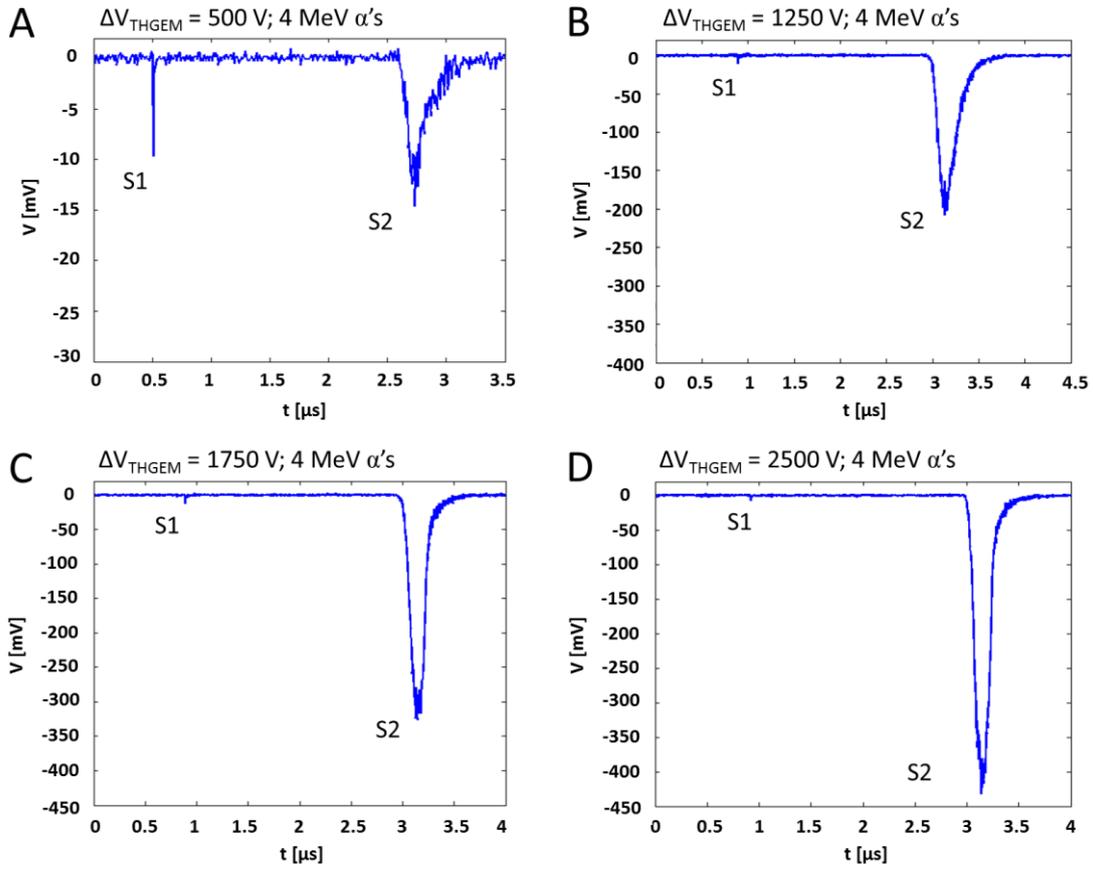

**Figure 6:** Samples of S1+S2 signal pairs induced in LXe by alpha particles in the setup of Figure 3B; the pulses were obtained with $E_{drift}$ = 1 kV/cm and THGEM voltages of: (A) 500 V, (B) 1250 V, (C) 1750 V and (D) 2500 V.

The dependence of the electron drift velocity on the drift field, based on the time difference between S1 and the initial rise of S2 (over the 5 mm drift gap), is shown in Figure 7. The results are in good agreement with published data [38].

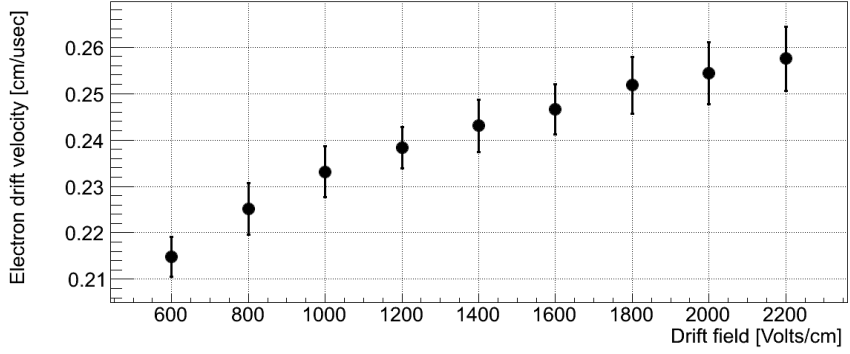

**Figure 7:** Electron drift velocity in LXe as a function of the drift field, measured with alpha-particle induced S1+S2 signal pairs (setup of Figure 3B). The error bars represent standard deviations (resulting from the spread of measured time difference between S1 and S2). $V_{THGEM}$ = 800 V, E = 0 below THGEM.

The spectra of 'raw' S2 pulse areas consisted of two components: a peak dominating the low-amplitude region with a monotonically decreasing tail extending to high amplitudes, and a Gaussian-like peak emerging at higher amplitudes (and becoming progressively more visible with increasing THGEM voltage). The first component resulted from events which included only S2 pulses with no S1; it could be completely removed by requiring that each recorded waveform should contain an S1 signal above noise, within a proper time window ~2 μs preceding S2 signals. Figure 8 shows the spectra, obtained before (A) and after (B) adding this requirement, for $V_{THGEM}$ = 2 kV and a $E_{drift}$ = 1 kV/cm. For comparison, the original energy spectrum of the $^{241}$Am source, measured with a solid-state detector, is shown as well (Figure 8C). The S2 spectrum shown in figure 8B is a sign of *proportional scintillation*. The relative FWHM of the distribution is 67% - considerably broader than the original alpha-particle energy spectrum shown in Figure 8C. This additional broadening could result from a convolution of several stochastic processes, namely the ionization and recombination along the alpha particle track, electron capture during their drift towards the THGEM, the efficiency of electron collection into the holes (in particular for events occurring in-between holes) and lastly – the scintillation process itself.

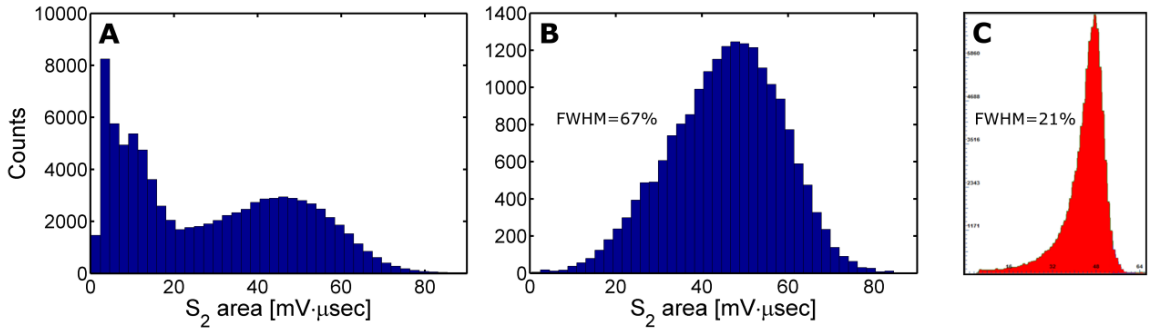

**Figure 8:** S2 electroluminescence spectra recorded in liquid xenon (T=173 K) from THGEM holes, with alpha particles in the setup of Figure 3B. (A) the raw spectrum, including events with no S1 pulses; (B) the same spectrum selecting S1 pulses with amplitudes > 6 mV; (C) the original non-spectroscopic alpha-particle spectrum, recorded with a solid-state detector. $V_{THGEM}$ = 2 kV; $E_{drift}$ = 1 kV/cm; $V_{PMT}$ = 850 V.

Events comprising S2 signals with no S1 could originate from: (1) 60 keV gammas emitted into the liquid, with the accompanying alpha particles emitted upward; (2) shallow alpha particle tracks, depositing most of their energy within the source itself and leaving a small amount of energy in the liquid, with an S1 signal buried in noise; (3) field emission of electrons from the top plate holding the source. The possibility of electron emissions from the THGEM itself was excluded because the rate of S2-only events was negligible with a zero drift field. Note that although the absorbed gammas leave only 60 keV in the liquid, the fraction of electrons which escape recombination in this case is ~80% at 1 kV/cm (compared to only ~3% for alphas) [1]; therefore, the number of electrons reaching the THGEM is only ~2.5 times smaller than for alpha particles.

The number of S2 photons reaching the PMT per electron passing through the THGEM hole was estimated by comparing the average S2 pulse-area to that of S1 (for the same PMT voltage) and using the estimated number of S1 photons reaching the PMT:

$$N_{photons}(S2\ at\ PMT) = N_{photons}(S1\ at\ PMT) \cdot \frac{S2\ average\ area}{S1\ average\ area}$$

As discussed earlier, the average S1 pulse area (0.06 mV·μs at PMT voltage of 850 V, see Figure 5), corresponds to ~500 photons reaching the PMT. Thus, for example, at a THGEM voltage of 1.75 kV, the average S2 area of 46 mV·μs (as taken from the appropriate histogram) corresponds to $\sim 500 \cdot 46/0.06 \approx 3.8 \times 10^5$ photons. The average energy needed to produce one electron-ion pair is 15.6 eV [1]. Thus, for an alpha particle of 4 MeV, 2.6×10[5] electron-ion pairs are created. Of these, only ~3% of the electrons escape recombination at a field of 1 kV/cm [1], i.e. ~7.7×10[3] electrons. Assuming all of these electrons reach the THGEM (without being captured) and are effectively focused into the THGEM holes, the number of photons reaching the PMT per electron passing through the hole is $\sim 3.8 \times 10^5 / 7.7 \times 10^3 \approx 50$. Assuming, for simplicity, that all photons are emitted at the center of the THGEM hole (where the field is maximal), the probability for an isotropically emitted photon to leave the hole towards the PMT was calculated to be 0.1 (for 0.3 mm diameter holes in a 0.4 mm thick electrode). Thus, the *total* estimated number of photons emitted over 4π per electron drifting through the hole is ~500 (for a THGEM voltage of 1.75 kV).

The dependence of the S2 pulse-area on the THGEM voltage for $E_{drift}$ = 1 kV/cm is shown in Figure 9, which also shows the estimated electroluminescence-photon yield (total number of photons emitted per electron drifting through the hole). As shown in the figure, the dependence is approximately linear over most of the range – as expected for proportional scintillation.

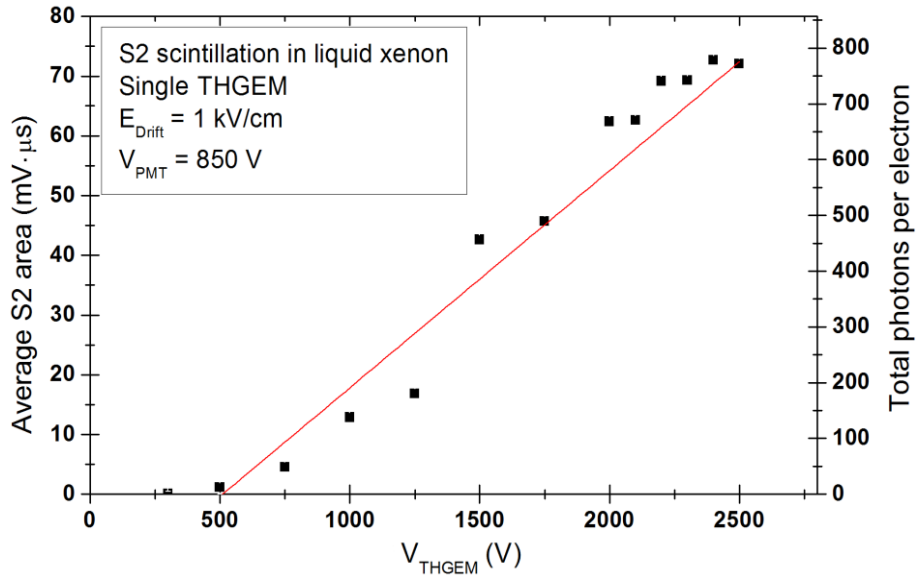

**Figure 9:** S2 average pulse-area and the estimated total number of photons per electron measured for alpha particles in liquid xenon (setup of Figure 3B), as a function of the THGEM voltage, for a drift field of 1 kV/cm and zero field below the THGEM.

## 4. Summary and discussion

In this work we demonstrated, for the first time, proportional electroluminescence in the holes of a THGEM electrode immersed in liquid xenon, using both 662 keV gammas and ~4 MeV alpha particles. This secondary scintillation process was found to depend linearly on the THGEM voltage.

With alpha particles, occasional faint S2 pulses began to appear at $V_{THGEM} \approx 300$ V becoming clearly distinct from the noise at $V_{THGEM} \approx 500$ V. For stable operation at $V_{THGEM} = 2$ kV and $E_{drift} = 1$ kV/cm we measured ~60 photons per drifting electron (after applying a linear fit to the entire data set), emitted from the hole in the forward direction. The estimated total photon yield emitted in these conditions within the hole over $4\pi$ is ~600 photons per drifting electron. At this voltage, the nominal field is $E_{hole} = 36$ kV/cm. The maximum stable THGEM voltage with gammas was 3.2 kV (2.6 kV with alpha particles); the estimated yield at $V_{THGEM} = 3.2$ kV is ~1000 photons per electron emitted over $4\pi$ (~100 emitted forward through the hole).

The spectrum of S2 signals obtained with alpha particles was Gaussian-like – indicating that the scintillation process is indeed proportional to the number of primary ionization electrons; its slight asymmetry was similar to that of the original non-spectroscopic alpha source. The spectrum was broad (~67% FWHM), indicating a large contribution of stochastic effects, including variations in electron recombination and capture, non-optimized focusing of the drifting electrons into the THGEM holes, hole-to-hole differences in wall roughness (affecting eventual reflections and resulting in local fields higher than the nominal one) and the scintillation process itself.

The S2 signals for alpha particles had a typical duration of 0.2-0.3 µs, in agreement with the time required for the drifting electrons to cross the holes. With gammas, the duration of S2 pulses was larger - up to ~1 µs - depending on the length and inclination of the recoiling electron tracks. A significant number of gamma events had a single S1 pulse followed by several S2 pulses, likely indicating multiple scattering. S2 peaks separated by a few hundred ns could be easily resolved. The electron drift velocity and its dependence on the drift field was measured with alpha particles and was found to be in very good agreement with published data. The THGEM electrode was stable throughout the measurements. Secondary effects (appearing as odd-shaped S2 signals with more than one peak) were very rare with alpha particles.

The present setup does not enable an exact determination of the electron lifetime, and thus the LXe purity is not well known. However, based on the width of the drift gap (5 mm) and large S2 signals, one can safely estimate that it is at least comparable to the drift time (2 µs), if not larger. Based on this value, we estimate that the concentration of impurities is in the ppm range, or smaller [37]. With this level of purity, it seems unlikely that the observed scintillation light from the THGEM holes originated from contaminants. We note that the data shown in this work was taken after two weeks of continuous Xe recirculation at 4-5 slpm through a state-of-the-art hot getter.

The measured S2 photon yield of about 500 photons/electron at a nominal THGEM field of 30 kV/cm is larger than, but of the same order of magnitude, as that observed in liquid argon for similar THGEM fields: ~500 photons/electron at 60 kV/cm [35]. In [35], the onset of observed electroluminescence was at much higher THGEM voltages (~8-9 kV compared to ~500 V here), but this can be attributed to the large dimensions of the electrode used (1.5 mm thickness and 1 mm hole diameter) and to the much smaller number of ionization electrons resulting from the absorption of 5.9 keV $^{55}$Fe x-ray photons used in that work (~30 times smaller, taking into account the different recombination rates along the track of an alpha particle and a photoelectron). The typical fields in which significant S2 signals were observed in this work are considerably lower than those observed in works on thin (few-µm diameter) wires [29-33]. The origin for this apparent discrepancy is presently not understood; it could be explained, in part, by the much longer path of the electrons drifting and inducing photons in the high field region of the THGEM hole.

The photon yield observed in this work is sufficiently high to consider further developments. One possible direction is the use of a single-stage THGEM as a means of generating S2 signals in single-phase liquid xenon TPCs. To be compatible with present-day S2 resolution in dual-phase detectors, the resolution of the THGEM-generated S2 signals should be further optimized; this requires a thorough study of the factors leading to the broadening of the spectrum reported in this work. The effective photon yield emitted from the holes may be enhanced by selecting other, highly reflective THGEM-substrate materials; e.g. Teflon (with proper metallic coating), whose reflectivity in LXe in the VUV region is close to unity [39]. An additional requirement from the THGEM substrate material is low natural radioactivity; this is met by Teflon, as well as by CIRLEX [40]. Another direction for further R&D is the development of cascaded, CsI-coated LHMs (liquid-hole-multipliers) discussed above, as suggested in [22]. Assuming a light yield of 80 photons per electron in the forward direction and 20% QE of CsI in LXe, each stage will have a light gain of ~10-15, amounting to the emission of thousands of photons in the forward direction by the last stage of a triple-stage LHM per photon or electron arriving at the first stage.

The preliminary promising results obtained in this work require further thorough investigation – with particular emphasis on their validity in ultra-high purity systems. Other THGEM geometries and materials, as well as GEMs and other hole-multipliers will be studied. The physical processes involved in the new concept of noble-liquid detectors with CsI-coated cascaded liquid hole-multipliers and their key features are already under study. They could have a major impact on future multi-ton rare-event experiments.

**Acknowledgments**

This work was supported in part by the Israel Science Foundation (Grant 477/10), the MINERVA Foundation (Project 710827) and by FCT and FEDER under the COMPETE program, through project CERN/FP/123614/2011. We would like to thank Prof. E. Aprile and Dr. R. Budnik of Columbia University, as well as Prof. Laura Baudis and Dr. Aaron Manalaysay of the University of Zurich for useful discussions on the design of the WILiX cryogenic system. We greatly acknowledge the technical assistance of Dr. S. Shchemelinin and of Mr. B. Pasmantirer, D. Front, M. Shoa, M. Klin and N. Priel of the Weizmann Institute. A.E.C. Coimbra greatly acknowledges the fruitful scientific discussions with Prof. J.M.F. dos Santos. A. Breskin is the W.P. Reuther Professor of Research in the Peaceful use of Atomic Energy.